
\input phyzzx
\font\elevenmib=cmmib10 scaled\magstephalf
\catcode`\@=11
\def\titlepage{\FRONTPAGE\papers\ifPhysRev\PH@SR@V\fi
 \leftline{\elevenmib Kanazawa}\vskip0.6cm
 \ifp@bblock\p@bblock \else\hrule height\z@ \rel@x \fi }
\catcode`\@=12

\pubnum={DPKU-9207}
\date={April 1992}
\titlepage
\title{Quantum Analysis of Jackiw and Teitelboim's Model
      \break for 1+1 D Gravity and Topological Gauge Theory}
\author{Haruhiko Terao
\footnote\dagger{E-mail address: AB0634@JPNKNZW1}
\footnote*{Reseach supported in part by SASAKAWA SCIENTIFIC RESEARCH GRANT of
THE JAPAN SCIENCE SOCIETY}
}
\vskip 0.5cm
\address{Department of Physics, Kanazawa University \break Kanazawa 920, Japan}

\abstract{
We study the BRST quantization of the 1+1 dimensional gravity model proposed by
Jackiw and Teitelboim
and also the topological gauge model which is equivalent to the gravity model
at least classically.
The gravity model quantized in the light-cone gauge is found to be a free
theory with a nilpotent BRST
charge. We show also that there exist twisted N=2 superconformal algebras in
the Jackiw-Teitelboim's
model as well as in the topological gauge model. We discuss the quantum
equivalence between the
gravity theory and the topological gauge theory. It is shown that these
theories are indeed
equivalent to each other in the light-cone gauge.
  }

\endpage

%
\chapter{Introduction}
It has been known for some time now that perturbative quantization fails in the
Einstein theory
for gravity due to non-renormalizability of the theory. On the other hand the
non-perturbative
canonical quantization is also quite difficult, because the constraint
equations, i.e.
Wheeler-DeWitt equations, are non-polynomial and highly complicated. Naive
consideration leads
us to these problems even for 2+1 dimensional pure gravity. However this is
somehow
surprising, since there is no local dynamical freedom in the 2+1 dimensional
pure gravity.

E. Witten addressed such a contradiction and showed the 2+1 dimensional pure
gravity in the Palatini
formalism may be formulated as the so-called Chern-Simons topological gauge
theory with a non-compact
gauge group.
\Ref\W{E. Witten, Nucl. Phys. B311 (1988/1989) 46.}
Namely the action is given by
$$
S={1 \over 2}\int d^3x \epsilon^{\mu\nu\rho}Tr \left(
A_{\mu}\partial_{\nu}A_{\rho}
+{2 \over 3} A_{\mu} A_{\nu} A_{\rho} \right),
\eqno(1.1)
$$
where the components of the gauge fields $A_\mu$ should be identified with
dribeins $e_\mu$ and
spin-connections $\omega_\mu$. Then this theory turns out to be renormalizable
as far as we
perform the perturbation around $e_\mu=\omega_\mu=0$, which is, however, a
singular
configulation as gravity. We may understand that such differences in the
quantum behavior
are caused by the different phase structures; namely the "unbroken phase" and
the "broken
phase", as Witten pointed out in the ref.[1]. Also in the canonical
quantization, the constraint
equations due to the gauge invariance are found to be polynomial ones, which
are actually shown to be
soluble, in sharp contrast to the Wheeler-DeWitt equations. Similar dramatic
simplification in the
canonical quantization has been discovered also for the 3+1 dimensional general
relativity.
\Ref\A{A. Ashtekar, New Perspectives in Canonical Gravity (Bibliopolis, Naples,
1988) and references
therein.}

Once we admit that the gauge theory of the Chern-Simons type describes the 2+1
dimensional gravity,
then the gravity seems to be very tractable and even soluble. However one may
note slight differences
in the ranges of the dynamical variables between those theories. In the
Einstein general relativity,
the inverse of the metric variable $g_{\mu\nu}$ appears, hence
$det(g_{\mu\nu})\equiv g$ cannot
vanishes. Also the volume element $\sqrt{-g}$ is demanded to be positive
definite from the
geometrical point of view. On the other hand the range of the gauge field is
unrestricted. Besides
the "volume" element which is given by $det(e_\mu^a)\equiv e$ in the
topological gauge theory is not
positive definite or the orientation of the space-time can be changed locally.
Strictly speaking the positive
definite volume element should be given by $|e|=\sqrt{-g}$ instead of $e$,
where we used the relation
$g_{\mu\nu}=e_\mu^a e_\nu^b \eta_{ab}$. \footnote{1)}{Our notation of the
Minkowski metric is
$\eta_{\mu\nu}=\eta_{ab}=diag(+1,-1,-1)$.} If we naively change the element $e$
to $|e|$, then the topological
features would disappear. Such differences in the ranges of the dynamical
variables seem to become more
important in quantum theories. It is because different quantum fluctuations are
expected to give rise to
different quantum theories generally, even if the classical trajectries are
same. However this problem seems to
be very subtle in the case of the 2+1 dimensional gravity, since there exists
no physical quantum fluctuation.
We may observe similar phenomena to appear in the Nambu-Goto string propagating
in two dimensional
space-time.
\Ref\FKT{K. Fujikawa, J. Kubo and H. Terao, Phys. Lett. B263 (1991) 371,
\nextline H. Terao and K. Yamada, Prog. Theor. Phys. 86 (1991) 1077.}

Here it may be natural to ask ourselves whether the Chern-Simons gauge theory
is really equivalent
to the 2+1 dimensional gravity in the quantum sense. One may also wonder if the
2+1 dimensional
gravity still has any topological features like the Chern-Simons gauge theory.
In order to explore
such questions, however, it seems to be necessary to find out first the
consistent quantum theory for
2+1 dimensional gravity. But this seems to be rather difficult to carry out.
Therefore in this paper
we would like to study 1+1 dimensional analogous models.

As is well known the Einstein action for pure gravity in two dimensional
space-time is trivial,
since it gives just the Euler number of the two dimensional surface. Polyakov
proposed to define the
gravitational theories in two dimensions as induced gravity by introducing
matter fields.
\Ref\Pa{A. M. Polyakov, Phys. Lett. 103B (1981) 207; Phys. Lett. (1981) 211.}
A lot of studies have been done on these models in various approaches. In the
continuum approaches the
light-cone gauge was found to be so powerful as to make it possible to solve
the models exactly.
\Ref\KPZ{A. M. Polyakov, Mod. Phys. Lett. A2 (1987) 893,
\nextline V. G. Knizhnik, A. M. Polyakov and A. B. Zamolodchikov, Mod. Phys.
Lett. A3 (1988) 819.}
\Ref\CR{A. H. Chamseddine and M. Reuter, Nucl. Phys. B317 (1989) 757.}
About a year later David, Distler and Kawai
\Ref\DDK{F. David, Mod. Phys. Lett. A3 (1988) 1651,
\nextline J. Distler and H. Kawai, Nucl. Phys. B321 (1989) 509.}
have succeeded in the quantization in the conformal gauge, which is familiar in
string theories.
These discoveries have been generating much progress in the continuum two
dimensional quantum
gravity theories.

Here, however, we are going to consider another model proposed by Jackiw and
Teitelboim
\Ref\JT{R. Jackiw, in: Quantum Theory of Gravity, ed. S. Christensen (Adam
Hilger, Bristol, 1984)
p.403,
\nextline C. Teitelboim, in: Quantum Theory of Gravity, ed. S. Christensen
(Adam Hilger, Bristol,
1984) p.327.}
\Ref\H{M. Henneaux, Phys. Rev. Lett. 54 (1985) 959.}
$$
S={1 \over 2}\int d^2x \sqrt{-g} (R+2\Lambda)\phi
\eqno(1.2)
$$
instead of the induced gravity theories. Because it has been found that this
action is actually
equivalent (at least classically) to the so-called $SO(2,1)$ topological gauge
theory,
\Ref\FK{T. Fukuyama and K. Kamimura, Phys. Lett. B169 (1985) 259.}
\Ref\TGT{K. Isler and C. A. Trugenberger, Phys. Rev. Lett. 63 (1989) 834,
\nextline A. H. Chamseddine and D. Wyler, Nucl. Phys. B340 (1990) 595.}
$$
S=\int d^2 \epsilon^{\mu\nu}Tr (\Phi F_{\mu\nu})
\eqno(1.3)
$$
as will be shown in detail later. We may see that in fact the actions (1.2) and
(1.3) are obtained
by dimensional reduction from the Einstein action and the Chern-Simons action
(1.1) in 2+1
dimensions. The questions that we are interested in here are the following. 1)
Are the gravitational
models defined by the actions (1.2) and (1.3) truly equivalent to each other in
the quantum
mechanical sense? 2) Are there any topological features appearing in the model
(1.2), though the
gauge theory (1.3) is indeed a so-called topological field theory? The answers
to these questions may
be useful to understand the relationship between the two kinds of formulations
for the 2+1
dimensional quantum gravity.

In section 2 we are going to perform the BRST quantization of Jackiw-
Teitelboim's model in the
light-cone gauge. We discuss also the quantization in the conformal gauge. The
$SO(2,1) (=SL(2,R))$
topological gauge theory is going to be quantized and it's topological algebra
will be given in
section 3. In section 4 we will consider the quantum equivalence between these
two models. Moreover,
in section 5, it will be shown that actually topological structure is realized
in Jackiw-Teitelboim's
model for 1+1 dimensional gravity as well as in the topological gauge theory.

\chapter{Quantization of Jackiw-Teitelboim's model}
First let us write down the starting action of Jackiw-Teitelboim's model again,
$$
S={1\over2}\int_{\Sigma}d^2x \sqrt{-g} (R+2\Lambda)\phi .
\eqno(2.1)
$$
Variation with respect to the scalar field $\phi$ generates a equation of
motion
$$
R+2\Lambda=0,
\eqno(2.2)
$$
which means that the curvature is fixed to a constant. Hereafter we would like
to assume that
$\Lambda >0$ and that the topology of the two dimensional surface $\Sigma$ is
fixed to $R\times
S^1$, which is the same topology as the de Sitter space.

The symmetric energy momentum tensor is given by the functional derivative
$$
T_{\mu\nu}=-{2 \over \sqrt{-g}}{\delta S \over \delta g^{\mu\nu}},
\eqno(2.3)
$$
which will play a important role in the quantum analysis later on. After some
straightforwards
calculations we may obtain
$$
T_{\mu\nu}=
\left( {1 \over 2}R\phi - \nabla^{\lambda}\nabla_{\lambda}\phi + \Lambda\phi
\right)g_{\mu\nu}
+\nabla_{\mu}\nabla_{\nu}\phi - R_{\mu\nu}\phi.
\eqno(2.4)
$$
The action is obviously invariant under the general coordinate transformation,,
$$
\eqalign{
\delta g_{\mu\nu}&=\nabla_{\mu}\xi_{\nu} +\nabla_{\nu}\xi_{\mu},  \cr
\delta \phi&= \xi^{\lambda}\partial_{\lambda}\phi.  \cr
}\eqno(2.5)
$$

2.1 Quantization in the light-cone gauge

We first perform the BRST quantization by using the light-cone
gauge,\refmark{\KPZ} which
are given by $g_{--}=0$ and $g_{-+}=1$. It will be convenient to define also
$g_{++}=2h_{++}$. Here $\pm$ denote the light-cone coordinates defined by
$x^{\pm}={1
\over \sqrt{2}}(x^0 \pm x^1)$. It should be noted here that $\sqrt{-g}$ is
fixed to be $1$ in this
gauge. The Ricci tensor and the scalar curvature are written down in terms of
$h_{++}$ as
$$
\eqalign{
R_{++}&=2h_{++}\partial_-^2  h_{++},\cr
R_{+-}&=R_{-+}=\partial_-^2 h_{++}, \cr
R_{--}&=0,\cr
R     &=2\partial_-^2 h_{++}. \cr}
\eqno(2.6)
$$
We may obtain the energy-momentum tensor in the light-cone gauge by
substituting the gauge
conditions into (2.4) as
$$
\eqalign{
T_{--}&=\partial_-^2 \phi,\cr
T_{+-}&=2h_{++}T_{--}+T_{+-}^N, \cr
T_{++}&=(2h_{++})^2 T_{--} + 2h_{++}T_{+-}^N + T_{++}^N,\cr
}\eqno(2.7)
$$
where we defined
$$
\eqalign{
T_{+-}^N&=-\left(\partial_+ - \partial_-h_{++}\right)\partial_-\phi
+\Lambda\phi  ,\cr
T_{++}^N&=\left( \partial_+ + \partial_-h_{++} - 2h_{++}\partial_-
\right)\partial_+\phi
          -\partial_+h_{++}\partial_-\phi.\cr
}\eqno(2.8)
$$
We note here that $T_{--}$ can be set to zero by using one of the equations of
motion, $\partial_-^2
\phi=0$. Therefore the new tensors may be expressed as
$$
\eqalign{
T_{+-}^N&=T_{+-}, \cr
T_{++}^N&=T_{++}-2h_{++}T_{+-}. \cr
}\eqno(2.9)
$$
Such structure has been already observed in the induced gravity
theory.\refmark{\CR}

Following the standard BRST procedure, the quantum action in this gauge will be
given as
$$
S=\int d^2x \bigg\{
\phi\left(\partial_-^2 h_{++} + \Lambda \right)
+2b_{++}\partial_- c^+ + b\left( \partial_+c^+ + \partial_-c^-
+2h_{++}\partial_-c^+ \right)
\bigg\},
\eqno(2.10)
$$
by taking account of the BRST transformations,
$$
\eqalign{
\delta^B g_{--}&= 2\partial_-c^+, \cr
\delta^B g_{+-}&=\partial_+c^+ + \partial_-c^- + 2h_{++}\partial_-c^+.  \cr
}\eqno(2.11)
$$
Here we introduced the ghost fields $c^+$ and $c^-$ for the general coordinate
transformation (2.5),
and also the anti-ghost fields $b_{++}$ and $b$. If we define a new anti-ghost
by $b'_{++}\equiv
2(b_{++} + h_{++}b)$, then the action (2.10) is reduced to a free one,
$$
S=\int d^2x \bigg\{
\phi\left(\partial_-^2 h_{++} + \Lambda \right)
+b'_{++}\partial_- c^+ + b\left( \partial_+c^+ + \partial_-c^- \right)
\bigg\}.
\eqno(2.12)
$$
This quantum action is actually invariant under the following BRST
transformations;
$$
\eqalign{
&\delta^B h_{++} =\partial_+c^- + c^+\partial_+h_{++} + c^-\partial_-h_{++} +
                  2\partial_+c^+h_{++},           \cr
&\delta^B \phi   =c^+\partial_+\phi + c^-\partial_-\phi,          \cr
&\delta^B c^+    =c^+\partial_+c^+ + c^-\partial_-c^+,            \cr
&\delta^B c^-    =c^+\partial_+c^- + c^-\partial_-c^-,            \cr
&\delta^B b'_{++}=T^N_{++} + c^+\partial_+b'_{++} + c^-\partial_-b'_{++} + 2
\partial_+c^+b'_{++},\cr
&\delta^B b      =T^N_{+-} + c^+\partial_+b + c^-\partial_-b.     \cr
}\eqno(2.13)
$$
Once we know these BRST transformations, then it is easy to find the
corresponding BRST currents
and they are found to be
$$
\eqalign{
J_{+}^{BRST}&=c^+\left( T^N_{++} + \partial_+c^+b'_{++} \right) +
              c^-\left( T^N_{+-} + c^+\partial_+b \right),      \cr
J_{-}^{BRST}&=c^+ T^N_{+-}.       \cr
}\eqno(2.14)
$$
We may easily show that they satisfy
$\partial_{-}J_{+}^{BRST}=\partial_{+}J_{-}^{BRST}=0$ by using
the equations of motion.

Let us mention here the residual symmetry after imposing the light-cone gauge.
The parameters for
the residual transformation should satisfy
$$
\eqalign{
\delta g_{--}&=2\partial_- \xi^+=0, \cr
\delta g_{+-}&=\partial_+ \xi^+ + \partial_- \xi^- +2h_{++}\partial_- \xi^+
=0.\cr
}\eqno(2.15)
$$
These equations are solved as
$$
\eqalign{
\xi^+&=\hat{\xi}^+(x^+), \cr
\xi^-&=\hat{\xi}^-(x^+)-x^-\partial_+\hat{\xi}^+(x^+). \cr
}\eqno(2.16)
$$
Therefore the transformation for the residual symmetry is given by the
parameters
$\hat{\xi}^+(x^+)$ and $\hat{\xi}^-(x^+)$. The corresponding conserved currents
$\hat{T}^{grav}_{+-}$
and $\hat{T}^{grav}_{++}$, which satisfy $\partial_-
\hat{T}^{grav}_{+-}=\partial_-
\hat{T}^{grav}_{++}=0$, are found to be
$$
\eqalign{
\hat{T}^{grav}_{+-}&= T^N_{+-}, \cr
\hat{T}^{grav}_{++}&= T^N_{++} + x^-\partial_+ T^N_{+-},  \cr
}\eqno(2.17)
$$
where $T^N_{+-}$ and $T^N_{++}$ are defined in (2.8).

Now we are in a position to perform the quantization of the action (2.12).
However it would be more
convenient to redefine the ghost variables and the anti-ghost variables as
$$
\eqalign{
\hat{c}^+    &\equiv c^+,\cr
\hat{c}^-    &\equiv c^- + x^-\partial_+c^+, \cr
\hat{b}_{++} &\equiv b'_{++} + x^-\partial_+b, \cr
\hat{b}      &\equiv b .\cr
}\eqno(2.18)
$$
Then the action (2.12) turns out to be a rather simple one;
$$
S=\int d^2x \bigg\{
\phi\left(\partial_-^2 h_{++} + \Lambda \right) +\hat{b}_{++}\partial_-
\hat{c}^+ +
\hat{b}\partial_-\hat{c}^-
\bigg\}.
\eqno(2.19)
$$
The BRST currents given in (2.14) are also rewritten in terms of the new ghost
variables into
$$
\eqalign{
J_+^{BRST} =& \hat{c}^+\hat{T}^{grav}_{++} +\hat{c}^-\hat{T}^{grav}_{+-}
 \cr
            &+ \hat{c}^+\partial_+\hat{c}^+\hat{b}_{++} +
\hat{c}^-\hat{c}^+\partial_+\hat{b}
            -\partial_+\left(x^-\hat{c}^+\hat{T}^{grav}_{+-} \right),
 \cr
J_-^{BRST} =&\hat{c}^+\hat{T}^{grav}_{+-},
        \cr
}\eqno(2.20)
$$
where the total divergence appearing in $J_+^{BRST}$ is irrelevant for the BRST
charge.

First let us consider the gravitational part of the action (2.19). The
equations of motion
$$
\eqalign{
\partial_-^2 h_{++} + \Lambda &=0,  \cr
\partial_-^2 \phi             &=0  \cr
}\eqno(2.21)
$$
are readily solved as
$$
\eqalign{
h_{++}(x^+,x^-) &=\hat{h}_{++}(x^+) + x^-\hat{h}_+(x^+) - {1 \over
2}(x^-)^2\Lambda,  \cr
\phi (x^+, x^-) &=\hat{\phi}(x^+) + x^-\hat{\phi}^+(x^+).
      \cr
}\eqno(2.22)
$$
On the other hand the equal time canonical commutation relations
$$
\eqalign{
\left[h_{++}(x^1),  \pi_h(y^1)\right]&=i\delta(x^1-y^1), \cr
\left[\phi(x^1),  \pi_{\phi}(y^1)\right]&=i\delta(x^1-y^1), \cr
}\eqno(2.23)
$$
where $\pi_h$ and $\pi_{\phi}$ are the canonical conjugate momentums of
$h_{++}$ and $\phi$
respectively, tell us that the component fields defined by (2.22) satisfy
$$
\eqalign{
\left[\hat{\phi}_+ (x^1),  \hat{h}_{++}(y^1)\right]&=i\delta(x^1-y^1), \cr
\left[\hat{h}_+(x^1),  \hat{\phi}(y^1)\right]&=i\delta(x^1-y^1). \cr
}\eqno(2.24)
$$
Since we have seen that these component fields are independent of $x^-$, the
operator
product expansions (O.P.E.'s) between these fields may be easily found to be
$$
\eqalign{
\hat{\phi}^+(x^+) \hat{h}_{++}(y^+) &\sim {1 \over x^+ - y^+},\cr
\hat{h}_+ (x^+) \hat{\phi}(y^+) &\sim {1 \over x^+ - y^+}, \cr
}\eqno(2.25)
$$
where we ignored the common irrelevant factor ${1 \over 2}$in the right-hand
sides. In a similar way the O.P.E.'s
between the ghost fields are given by
$$
\eqalign{
\hat{c}^+(x^+)\hat{b}_{++}(y^+) &\sim {1 \over x^+ - y^+},\cr
\hat{c}^-(x^+)\hat{b}(y^+)      &\sim {1 \over x^+ - y^+}. \cr
}\eqno(2.26)
$$

Now it is an important observation for the consistency to see whether the
quantum BRST charge is
nilpotent or not . For the purpose of verifying the BRST nilpotency, first we
should derive the
quantum algebra between the total "energy-momentum tensor" $\hat{T}^{tot}_{+-}$
and
$\hat{T}^{tot}_{++}$. The gravitational parts of these conserved currents are
given explicitly by
$$
\eqalign{
\hat{T}^{grav}_{+-}(x^+) &=\Lambda\hat{\phi} - \partial_+\hat{\phi}^+ +
\hat{\phi}^+\hat{h}_+,  \cr
\hat{T}^{grav}_{++}(x^+) &=
 \left( \partial_+^2 \hat{\phi} + \partial_+\hat{\phi}\hat{h}_+ \right)
-\left( \hat{\phi}^+\partial_+\hat{h}_{++} +
2\partial_+\hat{\phi}^+\hat{h}_{++} \right)  ,\cr
                         &\equiv \hat{T}^{c=2}_{++} + \hat{T}^{c=26}_{++}. \cr
}\eqno(2.27)
$$
Then the quantum algebra between these currents can be evaluated by using the
O.P.E.'s in (2.25) and
are found to be in O.P.E. forms
$$
\eqalign{
\hat{T}^{grav}_{+-}(x^+)\hat{T}^{grav}_{+-}(y^+) &\sim 0,  \cr
\hat{T}^{grav}_{++}(x^+)\hat{T}^{grav}_{++}(y^+) &\sim
{(2+26)/2 \over (x^+ - y^+)^4} + {2\hat{T}^{grav}_{++}(y^+) \over (x^+ -
y^+)^2} +
{\partial_+\hat{T}^{grav}_{++}(y^+) \over x^+ - y^+},                  \cr
\hat{T}^{grav}_{++}(x^+)\hat{T}^{grav}_{+-}(y^+) &\sim
{\partial_+\hat{T}^{grav}_{+-}(y^+) \over x^+ - y^+}.                  \cr
}\eqno(2.28)
$$
Therefore we see that the Virasoro algebra of the gravitational part carries
the central charge of
$28$ and that $\hat{T}^{grav}_{+-}$ is a commuting current with spin $0$. The
ghost parts of the
"energy-momentum tensor" from the action (2.19) can be derived as the conserved
currents for the
residual symmetry, and are given by
$$
\eqalign{
\hat{T}^{gh}_{+-}(x^+) &=\hat{c}^+\partial_+\hat{b},                    \cr
\hat{T}^{gh}_{++}(x^+) &=-\hat{c}^-\partial_+\hat{b} +
                         \left(\hat{c}^+\partial_+\hat{b}_{++} +
                         2\partial_+\hat{c}^+\hat{b}_{++} \right) ,     \cr
                       &=\hat{T}^{c=-2}_{++} + \hat{T}^{c=-26}_{++} .   \cr
}\eqno(2.29)
$$
By using the O.P.E.'s (2.26) we may see $\hat{T}^{gh}_{+-}$ and
$\hat{T}^{gh}_{++}$
 satisfy the similar algebra to (2.27);
$$
\eqalign{
\hat{T}^{gh}_{+-}(x^+)\hat{T}^{gh}_{+-}(y^+) &\sim 0,  \cr
\hat{T}^{gh}_{++}(x^+)\hat{T}^{gh}_{++}(y^+) &\sim
{-(2+26)/2 \over (x^+ - y^+)^4} + {2\hat{T}^{gh}_{++}(y^+) \over (x^+ - y^+)^2}
+
{\partial_+\hat{T}^{gh}_{++}(y^+) \over x^+ - y^+},                  \cr
\hat{T}^{gh}_{++}(x^+)\hat{T}^{gh}_{+-}(y^+) &\sim
{\partial_+\hat{T}^{gh}_{+-}(y^+) \over x^+ - y^+},                  \cr
}\eqno(2.30)
$$
from which the contribution of the ghost part to the central charge is read off
to be $-28$.
Therefore there appears no anomaly in the Virasoro algebra of the total
energy-momentum tensor
$\hat{T}^{tot}=\hat{T}^{grav}+\hat{T}^{gh}$;
$$
\hat{T}^{tot}_{++}(x^+)\hat{T}^{tot}_{++}(y^+) \sim
{2\hat{T}^{tot}_{++}(y^+) \over (x^+ - y^+)^2} +
{\partial_+\hat{T}^{tot}_{++}(y^+) \over x^+ - y^+},
\eqno(2.31)
$$
Consequently we can indeed verify the nilpotency of the BRST charge defined by
(2.19);
$$
\left( Q_+ ^{BRST} \right)^2 =\left( Q_- ^{BRST} \right)^2 = 0.
\eqno(2.32)
$$
Actually this may be expected before the calculations, because the anomaly does
not seem to appear
without any physical freedoms in the local dynamics.

2.2 Quantization in the conformal gauge

The conformal gauge, which is very familiar to string physicists, is defined by
$g_{\mu\nu}=e^{\varphi}\eta_{\mu\nu}$ or
$$
\eqalign{
g_{++} &= g_{--} = 0, \cr
g_{+-} &= g_{-+} = e^{\varphi}, \cr
}\eqno(2.33)
$$
where we note that the volume form $\sqrt{-g} = e^{\varphi}$ is kept positive
definite without
restricting the range of the conformal mode $\varphi$. The good point of this
gauge is that the
so-called conformal symmetry, which includes the global Lorentz symmetry, is
maintained as the
residual symmetry. However this gauge will bring some troubles to proceed the
quantization,
especially in the case of non-zero cosmological constant, as is seen later on.

By inserting the gauge conditions (2.33) into (2.4) the energy-momentum tensor
becomes
$$
\eqalign{
T_{++}^{grav} &=\nabla_+\nabla_+\phi = \partial_+^2\phi -
\partial_+\varphi\partial_+\phi,  \cr
T_{--}^{grav} &=\nabla_-\nabla_-\phi = \partial_-^2\phi -
\partial_-\varphi\partial_-\phi,  \cr
T_{+-}^{grav} &=-\partial_+\partial_-\phi + \Lambda \phi e^{\varphi}.
                \cr
}\eqno(2.34)
$$
The quantum action in the conformal gauge also is obtained through the usual
BRST procedure and
$$
S=\int d^2x \bigg\{
\phi(-\partial_+\partial_-\varphi + \Lambda  e^{\varphi}) +
b_{++}\partial_-c^+ + b_{--}\partial_+c^-
\bigg\}.
\eqno(2.35)
$$
It should be noted that there is an interaction term unless $\Lambda=0$ in
contrast to the action
(2.12) in the light-cone gauge. The equations of motion of the gravitational
part are readily derived
as
$$
\eqalign{
\partial_{+}\partial_{-}\varphi &= \Lambda e^{\varphi}      ,  \cr
\partial_{+}\partial_{-}\phi    &= \Lambda \phi e^{\varphi} .  \cr
}\eqno(2.36)
$$
The first equation is the so-called Liouville equation. The second one means
$T^{grav}_{+-}=0$, namely
the presence of the conformal symmetry in the classical level. It is also a
rather easy task to find
the BRST transformations and they are given by
$$
\eqalign{
&\delta^B \varphi =\partial_+c^+ + \partial_-c^- + c^+\partial_+\varphi +
c^-\partial_-\varphi,  \cr
&\delta^B \phi    =c^+\partial_+\phi + c^-\partial_-\phi,          \cr
&\delta^B c^+     =c^+\partial_+c^+ + c^-\partial_-c^+,            \cr
&\delta^B c^-     =c^+\partial_+c^- + c^-\partial_-c^-,            \cr
&\delta^B b_{++}  =T_{++}^{grav} + c^+ \partial_+ b_{++} + 2\partial_+ c^+
b_{++} +
                   c^- \partial_- b_{++},                          \cr
&\delta^B b_{--}  =T_{--}^{grav} + c^- \partial_- b_{--} + 2\partial_- c^-
b_{--} +
                   c^+ \partial_+ b_{--}.                          \cr
}\eqno(2.37)
$$
By the Noether method using (2.37) we may derive the BRST currents;
$$
\eqalign{
J_+ ^{BRST} &= c^+ \left( T_{++}^{grav} + {1 \over 2}T_{++}^{gh} \right),  \cr
J_- ^{BRST} &= c^- \left( T_{--}^{grav} + {1 \over 2}T_{--}^{gh} \right),  \cr
}\eqno(2.38)
$$
where we introduced the energy-momentum tensor of the ghost part,
$$
\eqalign{
T_{++}^{gh} &= c^+ \partial_+ b_{++} + 2\partial_+ c^+ b_{++}, \cr
T_{--}^{gh} &= c^- \partial_- b_{--} + 2\partial_- c^- b_{--}. \cr
}\eqno(2.39)
$$
Here we note the structure of the BRST currents are identical to one of the
induced gravity or
string theories in the conformal gauge. We can verify easily also the
conservation of these BRST
currents, $\partial_- J_+ ^{BRST} =\partial_+ J_- ^{BRST} =0$, by using the
equations of motion
(2.36). The total energy-momentum tensor is given by
$$
\eqalign{
T_{++}^{tot} &= T_{++}^{grav} + T_{++}^{gh}, \cr
T_{--}^{tot} &= T_{--}^{grav} + T_{--}^{gh}, \cr
}\eqno(2.40)
$$
which are the generators of the conformal symmetry.

Now we would like to consider to quantize the action (2.35). As is seen from
the equations of motion
(2.36), this problem may be closely related to the quantization of the
Liouville theory.
\Ref\GN{J. L. Gervais and A. Neveu, Nucl. Phys. B199 (1982) 59; B209 (1982)
125; B224 (1983) 329;
B238 (1984) 125; B238 (1884) 396.}
\Ref\L{T. L. Curtright and C. B. Thorn, Phys. Rev. Lett. 48 (1982) 1309,
\nextline E. D'Hooker and R. Jackiw, Phys. Rev. D26 (1982) 3517; Phys. Rev.
Lett. 50 (1983) 1719,
\nextline E. D'Hooker, D. Z. Freedman and R. Jackiw, Phys Rev. D28 (1983) 2583,
\nextline T. Yoneya, Phys. Lett. B148 (1984) 111.}
{}From the action
$$
\eqalign{
S=\int d^2x \bigg\{
&{1 \over 2} (\partial_0 + \partial_1 )\phi (\partial_0 - \partial_1 )\varphi +
 {1 \over 2} (\partial_0 - \partial_1 )\phi (\partial_0 + \partial_1 )\varphi +
2\Lambda \phi e^{\varphi}                                               \cr
&+b_{++}(\partial_0 - \partial_1)c^+ +
b_{--}(\partial_0 + \partial_1)c^- \bigg\},																											\cr
}\eqno(2.41)
$$
where we rescaled the variables slightly for the simplicity, the canonical
conjugate momentums are
given by $$
\eqalign{
\pi_{\phi} &= \partial_0 \varphi, \cr
\pi_{\varphi} &= \partial_0 \phi, \cr
\pi_{c^+} &= b_{++}, \cr
\pi_{c^-} &= b_{--}. \cr
}\eqno(2.42)
$$
On the other hand the energy-momentum tensor of the gravitational sector may be
expressed in terms
of these canonical variables as
$$
\eqalign{
T_{++}^{grav} &= -{1\over 2}:(\pi_{\phi} + \partial_1 \varphi)(\pi_{\varphi} +
\partial_1 \phi):
+ \Lambda \phi e^{\varphi} + \partial_1 (\pi_{\varphi} + \partial_1 \phi), \cr
T_{--}^{grav} &= -{1\over 2}:(\pi_{\phi} - \partial_1 \varphi)(\pi_{\varphi} -
\partial_1 \phi):
+ \Lambda \phi e^{\varphi} - \partial_1 (\pi_{\varphi} - \partial_1 \phi), \cr
}\eqno(2.43)
$$
where $::$ denotes the normal ordering. By using the equal time commutation
relations,
$$
\eqalign{
\left[\phi(x^1),  \pi_{\phi}(y^1)\right]&=i\delta(x^1-y^1), \cr
\left[\varphi(x^1),  \pi_{\varphi}(y^1)\right]&=i\delta(x^1-y^1), \cr
}\eqno(2.44)
$$
we may extract the short distance singularities from the product of two
$T_{++}^{grav}
(x^1)$'s. After some manipulations\refmark{\GN,\L} they are found to be
$$
\eqalign{
T_{++}^{grav}(x^1)T_{++}^{grav}(y^1) \sim
&{1 \over (x^1 - y^1)^4} + {2T^{grav}_{++}(y^1) \over (x^1 - y^1)^2} +
 {\partial_1 T^{grav}_{++}(y^1)  \over x^1 - y^1}                          \cr
&- \Lambda \left[ {1 \over (x^1 - y^1)^2}e^{\varphi} +
{1/2 \over x^1 - y^1}\partial_1\varphi e^{\varphi} \right].                 \cr
}\eqno(2.45)
$$
Similarly we obtain
$$
T_{++}^{gh}(x^1)T_{++}^{gh}(y^1) \sim
{-26/2 \over (x^1 - y^1)^4} + {2T^{gh}_{++}(y^1) \over (x^1 - y^1)^2} +
{\partial_1 T^{gh}_{++}(y^1) \over x^1 - y^1}
\eqno(2.46)
$$
for the energy-momentum tensor of the ghosts. Thus it seems that the anomalies
do not cancel each
other between the gravitational part and the ghost part, even if $\Lambda=0$,
contrary to the results
in the light-cone gauge (2.31). In the presence of the non-zero cosmological
constant $\Lambda$, the
situation looks much worse. Indeed the short distance singularities
proportional to $\Lambda$ in
(2.45) seem to prevent the Virasoro algebra from closing.

These anomalous results, however, are caused by the inconsistency of our
quantization procedure.
Actually it is necessary to take care of the renormalization effects due to the
non-trivial
interactions. Besides we have to evaluate the Jacobian which is probably
generated through the
change of the dynamical variable $g_{+-}=e^{\varphi}$ to $\varphi$. Suppose,
therefore, the
energy-momentum tensor is improved to
$$
\eqalign{
T_{++}^{grav} = &-{1\over 2}:(\pi_{\phi} + \partial_1 \varphi)(\pi_{\varphi} +
\partial_1 \phi):
                 + \Lambda^{ren} \phi e^{\alpha\varphi}                 \cr
                &+ \partial_1 (\pi_{\varphi} + \partial_1 \phi)
                 + \beta \partial_1 (\pi_{\phi} + \partial_1 \varphi) , \cr
}\eqno(2.47)
$$
where $\alpha$ and $\beta$ are unknown parameters to be determined by the
consistency.
\refmark{\L,\DDK} If we set the parameters to
$$
\eqalign{
\alpha &= 1, \cr
\beta  &= {1 \over 2},  \cr
}\eqno(2.48)
$$
then the total energy-momentum tensor $T_{++}^{tot}$ indeed satisfies the
anomaly free and closed
algebra;
$$
T_{++}^{tot}(x^1)T_{++}^{tot}(y^1) \sim
{2 \over (x^1 - y^1)^2}T^{tot}_{++}(y^1) + {1 \over x^1 -
y^1}\partial_1T^{tot}_{++}(y^1).
\eqno(2.49)
$$
The $(-)$ sector also enjoys the similar improvement. Thus Jackiw-Teitelboim's
model may be
quantized consistently also in the conformal gauge, if we use the
energy-momentum tensor given by
(2.47) and (2.48). However further investigations are needed to see whether the
quantization using
(2.47) really gives us the same results as the light-cone quantization. The
quantization of the similar action
has been examined by the perturbative approach also.
\Ref\AN{M. Abe and N. Nakanashi, preprint RIM-847.}

\chapter{Quantization of the $SL(2,R)$ topological gauge theory}

In this section we are going to perform the BRST quantization of the
topological gauge theory with
$SO(2,1)$ gauge group;
$$
S=\int d^2 x \epsilon^{\mu\nu}Tr(\Phi F_{\mu\nu}),
\eqno(3.1)
$$
where
$F_{\mu\nu}=\partial_{\mu}A_{\nu}-\partial_{\nu}A_{\mu}+[A_{\mu},A_{\nu}]$
is a field strength. The $SO(2.1)$ or 2d de Sitter algebra is
$$
\eqalign{
\left[ P_a, P_b \right] &= \epsilon_{ab} J  ,\cr
\left[ J, P_a \right]   &= \epsilon_{ab} P^b  ,\cr
}\eqno(3.2)
$$
where $a$ and $b$ are $0$ or $1$. For the later conveniences let us define new
generators by
$T_{\pm} \equiv {1 \over \sqrt{2}}(P_0 \pm P_1)$ and $T_0 \equiv J$. Then they
form a $SL(2,R)$
algebra,
$$
\eqalign{
\left[ T_0, T_{\pm} \right] &= \pm T_{\pm}  ,\cr
\left[ T_+, T_- \right]   &= T_0  .\cr
}\eqno(3.3)
$$
If we expand the gauge field $A_{\mu}$ and the scalar field $\phi$ with respect
to the $SL(2,R)$
generators as
$$
\eqalign{
A_{\mu} &= \sqrt{\Lambda}e_{\mu}^+ T_+ +\sqrt{\Lambda}e_{\mu}^- T_-
+\omega_{\mu} T_0, \cr
\Phi    &= {1 \over \sqrt{\Lambda}}\phi^+ T_+ +{1 \over \sqrt{\Lambda}}\phi^-
T_- +\phi T_0, \cr
}\eqno(3.4)
$$
then the action in terms of the component fields turns to be
$$
\eqalign{
S=\int d^2 x \bigg\{
& \phi^- \left( D_+ e_-^+ - D_- e_+^+ \right)
 +\phi^+ \left( D_+ e_-^- - D_- e_+^- \right)          \cr
&+\phi   \left( \partial_+ \omega_- - \partial_- \omega_+ - \Lambda(e_+^+ e_-^-
+ e_+^-e_-^+) \right)
\bigg\},   \cr
}\eqno(3.5)
$$
where $D_{\pm}$ denote the covariant derivatives defined by
$D_{\mu}e^{\pm}_{\nu}=(\partial_{\mu} \pm \omega_{\mu})e^{\pm}_{\nu}$.
The multipliers $\phi^{\pm}$ generate the so-called torsion-free conditions;
$D_+ e_-^{\pm}- D_- e_+^{\pm}=0$. If we solve the spin connections
$\omega_{\pm}$ in
terms of the zweibeins $e_{\pm}^{\pm}$ by using these torsion-free conditions,
then indeed the
action (3.5) is found to be reduced to the gravitational action
(2.1).\refmark{\TGT} However we need
to change $e=e_+^+ e_-^- - e_+^-e_-^+$ to $\sqrt{-g}=|e|$ to obtain (2.1). This
ambiguity of the sign
does not affect the equations of motion, hence we may say that this topological
gauge theory is
equivalent to Jackiw-Teitelboim's model in the classical level. Hereafter let
us rescale
$e_{\mu}^{\pm}$ and $\phi^{\pm}$ to absorb the factor $\sqrt{\Lambda}$.
 \footnote{2)}
{In the case of $\Lambda <0$ the gauge group should be $SO(1,2)$ (anti-de
Sitter group)
instead of $SO(2,1)$. If $\Lambda=0$, then we have to define the theory in the
limit of $\Lambda
\to 0$. Such cases are also discussed in the ref.[\TGT], though we are not
going to mention in this
paper.}

The action (3.1) is obviously invariant under the $SL(2,R)$ gauge
transformations which are
$$
\eqalign{
\delta A_{\mu} &= \partial_{\mu}\epsilon + \left[ A_{\mu}, \epsilon \right],
\cr
\delta \Phi    &= \left[ \Phi, \epsilon\right],  \cr
}\eqno(3.6)
$$
where we introduced the gauge parameter $\epsilon = \epsilon^+T^- +
\epsilon^-T^+ + \epsilon^0T^0$.
One may note that this action is invariant also under the general coordinate
transformations. In
practice, however, we do not have to concern this symmetry, because the general
coordinate
transformations and the gauge transformations are "reducible" to each other.
Actually we will see
that it is enough to fix only the gauge transformations (3.6) in order to
construct the quantum
action.

3.1 Quantization in $A_- ^a =0$ gauge

We would like to consider first the BRST quantization by the gauge conditions
$A_- ^a$ or
$$
e_- ^+ = e_- ^- = \omega _- =0.
\eqno(3.7)
$$
 Other gauges also will be examined in the next section. We note that the gauge
conditions
(3.7) give a vanishing "volume form" $e=e_+ ^+ e_- ^- - e_+ ^- e_- ^+ =0$,
which is a singular
configuration as gravity. By introducing ghosts $\tilde{c}$'s and anti-ghosts
$\tilde{b}$'s, the
BRST invariant quantum action in the $A_- ^a=0$ gauge is given by
$$
\eqalign{
S=\int d^2x \bigg\{
& -\phi^-\partial_-e_+^+ - \phi^+\partial_-e_+^- - \phi\partial_-\omega_+  \cr
&+\tilde{b}_+^-\partial_-\tilde{c}^+ + \tilde{b}_+^+\partial_-\tilde{c}^- +
  \tilde{b}_+  \partial_-\tilde{c}
\bigg\}.
}\eqno(3.8)
$$
The BRST transformations are easily found to be the following,
$$
\eqalign{
&\delta^B e_+ ^+           =\partial_+\tilde{c}^+ + \omega_+\tilde{c}^+ -
e_+^+\tilde{c},\cr
&\delta^B e_+ ^-           =\partial_+\tilde{c}^- - \omega_+\tilde{c}^- +
e_+^-\tilde{c},\cr
&\delta^B \omega_+         =\partial_+\tilde{c} + e_+^+\tilde{c}^- -
e_+^-\tilde{c}^+,   \cr
&\delta^B \phi^+           =\phi \tilde{c}^+ - \phi^+\tilde{c},
         \cr
&\delta^B \phi^-           =\phi^- \tilde{c} - \phi\tilde{c}^-,
         \cr
&\delta^B \phi             =\phi^+ \tilde{c}^- - \phi^-\tilde{c}^+,
         \cr
&\delta^B \tilde{c}^+      =\tilde{c}^+\tilde{c}  ,
									\cr
&\delta^B \tilde{c}^-      =\tilde{c}\tilde{c}^-,
									\cr
&\delta^B \tilde{c}        =\tilde{c}^-\tilde{c}^+,
									\cr
&\delta^B \tilde{b}_+^{\pm}={J^{tot}}_+ ^{\pm} ,
         \cr
&\delta^B \tilde{b}_+      ={J^{tot}}_+ ^0 .
         \cr
}\eqno(3.9)
$$
Here ${J^{tot}}_+ ^a \equiv {J^{gauge}}_+ ^a+ {J^{gh}}_+ ^a$ are the conserved
currents for the
residual $SL(2,R)$ symmetry, which are given explicitly by
$$
\eqalign{
{J^{gauge}}_+ ^+ &=-(\partial_+ + \omega_+)\phi^+ + e_+^+\phi ,            \cr
{J^{gauge}}_+ ^- &=-(\partial_+ - \omega_+)\phi^- - e_+^-\phi ,            \cr
{J^{gauge}}_+ ^0 &=-\partial_+\phi + e_+^-\phi^+ - e_+^+\phi^- ,           \cr
{J^{gh}}_+ ^+    &=\tilde{b}_+^+\tilde{c} - \tilde{b}_+  \tilde{c}^+ - ,   \cr
{J^{gh}}_+ ^-    &=\tilde{b}_+  \tilde{c}^- - \tilde{b}_+^-\tilde{c}   - , \cr
{J^{gh}}_+ ^0    &=\tilde{b}_+^-\tilde{c}^+ - \tilde{b}_+^+\tilde{c}^- - . \cr
}\eqno(3.10)
$$
Note that there are no symmetry currents in the $(-)$ sector. In terms of these
currents the BRST
current may be expressed neatly as
$$
\eqalign{
{J^{BRST}} _+ = & \tilde{c}^+ \left( {J^{gauge}}_+ ^- +{1 \over 2}{J^{gh}}_+ ^-
\right)  \cr
                &+\tilde{c}^- \left( {J^{gauge}}_+ ^+ +{1 \over 2}{J^{gh}}_+ ^+
\right)  \cr
                &+\tilde{c}   \left( {J^{gauge}}_+ ^0 +{1 \over 2}{J^{gh}}_+ ^0
\right). \cr
}\eqno(3.11)
$$

In this case the quantization is rather simple. The non-trivial O.P.E.'s
between the fields are also
readily found to be
$$
\eqalign{
e_+ ^{\pm}(x^+) \phi^{\mp}(y^+)             &\sim {1 \over x^+ - y^+}, \cr
\omega_+ (x^+) \phi (y^+)                   &\sim {1 \over x^+ - y^+}, \cr
\tilde{c}^{\pm}(x^+) \tilde{b}_+^{\mp}(y^+) &\sim {1 \over x^+ - y^+}, \cr
\tilde{c}      (x^+) \tilde{b}_+      (y^+) &\sim {1 \over x^+ - y^+}. \cr
}\eqno(3.12)
$$
By using these O.P.E.'s (3.12) we may easily see that the currents
$J^{gauge}_+$ and $J^{gh}_+$
indeed form the following $SL(2.R)$ Kac-Moody algebras;
$$
\eqalign{
{J^{gauge}}_+^+(x^+) {J^{gauge}}_+^-(y^+)
&\sim {-2 \over (x^+ - y^+)^2} + {{J^{gauge}}_+^0(y^+) \over x^+ - y^+},  \cr
{J^{gauge}}_+^0(x^+) {J^{gauge}}_+^0(y^+)
&\sim {-2 \over (x^+ - y^+)^2}  ,  \cr
{J^{gauge}}_+^0(x^+) {J^{gauge}}_+^{\pm}(y^+)
&\sim \pm {{J^{gauge}}_+^{\pm}(y^+) \over x^+ - y^+}  \cr
}\eqno(3.13)
$$
and
$$
\eqalign{
{J^{gh}}_+^+(x^+) {J^{gh}}_+^-(y^+)
&\sim {2 \over (x^+ - y^+)^2} + {{J^{gh}}_+^0(y^+) \over x^+ - y^+},  \cr
{J^{gh}}_+^0(x^+) {J^{gh}}_+^0(y^+)
&\sim {2 \over (x^+ - y^+)^2}  ,  \cr
{J^{gh}}_+^0(x^+) {J^{gh}}_+^{\pm}(y^+)
&\sim \pm {{J^{gh}}_+^{\pm}(y^+) \over x^+ - y^+}.  \cr
}\eqno(3.14)
$$
Here we note that the Schwinger terms in the gauge sector and the ghost sector
cancel each other.
Thus it is verified that there appears no anomaly in the Kac-Moody algebra of
the total current
${J^{tot}}_+$, as is expected. The nilpotency of the BRST charge
$(Q^{BRST})^2=0$ follows from
these calculations immediately.

In the last part of this section we would like to discuss the topological
algebra, which is expected
to exist in this topological gauge theory. It is known that two dimensional
topological field
theories may be characterized commonly by the twisted $N=2$ superconformal
algebras (SCA's).
\Ref\EY{T. Eguchi and S.K. Yang, Mod. Phys. Lett. A5 (1990) 1693.}
Indeed it is found that this topological gauge model also has a twisted $N=2$
SCA. The set of
generators of the twisted $N=2$ SCA consists of the energy-momentum tensor
$T_{++}$, a spin $1$
super-current $G_+$, a spin $2$ super-current $\bar{G}_{++}$ and a $U(1)$
current $I_+$. If we
define them as
$$
\eqalign{
&T_{++}      \equiv  \partial_+\phi^+ e_+^- + \partial_+\phi^- e_+^+ +
\partial_+\phi\omega_+
                    +\partial_+\tilde{c}^+ \tilde{b}_+^- +
\partial_+\tilde{c}^- \tilde{b}_+^+
                    +\partial_+\tilde{c}   \tilde{b}_+,
               \cr
&G_+         \equiv {J^{BRST}}_+ ,
															\cr
&\bar{G}_{++}\equiv \tilde{b}_+^+e_+^- + \tilde{b}_+^-e_+^+ + \tilde{b}_+
\omega_+,            \cr
&I_+         \equiv {J^{gh}}_+
                   =-\tilde{b}_+^+\tilde{c}^- -\tilde{b}_+^-\tilde{c}^+
-\tilde{b}_+\tilde{c} ,\cr
}\eqno(3.15)
$$
where ${J^{BRST}}_+$ is the BRST current given in (3.11) and ${J^{gh}}_+$ is
the ghost number
current,
\footnote{3)}{
It should be noted that the energy-momentum tensor $T_{++}$ in (3.15) is not
the one by
Sugawara construction from the Kac-Moody currents. Such topological models also
have been
considered.
\Ref\KM{S. Komata and K. Mohri, preprint UT-Komaba 91-11 revised (1991).}
}
then these currents are found to form the following algebra;
$$
\eqalign{
&T_{++}(x^+) T_{++}(y^+)       \sim
{2T_{++}(y^+) \over (x^+ - y^+)^2} +{\partial_+T_{++}(y^+) \over x^+ - y^+},
  \cr
&T_{++}(x^+) G_{+}(y^+)        \sim
{G_+ (y^+) \over (x^+ - y^+)^2} +{\partial_+ G_+ (y^+) \over x^+ - y^+},
\cr
&T_{++}(x^+) \bar{G}_{++}(y^+) \sim
{\bar{G}_{++} (y^+) \over (x^+ - y^+)^2} +{\partial_+ \bar{G}_{++} (y^+) \over
x^+ - y^+},      \cr
&T_{++}(x^+) I_{+}(y^+)        \sim
{-3 \over (x^+ - y^+)^3} + {I_+(y^+) \over (x^+ - y^+)^2} + {\partial_+I_+(y^+)
\over x^+ - y^+},\cr
&I_{+}(x^+)  G_{+}(y^+)        \sim {G_+(y^+) \over x^+ - y^+}  ,\cr
&I_{+}(x^+) \bar{G}_{++}(y^+)  \sim {-\bar{G}_{++}(y^+) \over x^+ - y^+},\cr
&I_{+}(x^+)  I_{+}(y^+)        \sim {3 \over (x^+ - y^+)^2} ,           \cr
&G_{+}(x^+) \bar{G}_{++}(y^+)  \sim
{3 \over (x^+ - y^+)^3} + {I_+(y^+) \over (x^+ - y^+)^2} + {T_{++}(y^+) \over
x^+ - y^+},   \cr
&G_{+}(x^+) G_{+}(y^+)         \sim  \bar{G}_{++}(x^+) \bar{G}_{++}(y^+) \sim
0.
}\eqno(3.16)
$$
This is nothing but the $N=2$ twisted SCA with the central charge $\hat{c}=3$.
Now it would be
natural to wonder if such a twisted SCA is realized also in Jackiw-Teitelboim's
model which was
examined extensively in the last section. This will be considered in section 5.

\chapter{On the quantum equivalence}

In the last section we have seen that both of Jackiw-Teitelboim's model and the
$SL(2,R)$
topological gauge model can be consistently quantized in the BRST formulation.
What we are
interested in now is the quantum equivalence between those two models. In order
to see the
equivalence it would be one way to examine the physical spaces which are
determined by the
BRST-cohomologies. However the BRST currents obtained in (2.20) and in (3.11)
look rather different in
the structure from each other. The BRST current for Jackiw-Teitelboim's model
is based on the
energy-momentum tensor $\hat{T}^{grav}_{++}$ and the $U(1)$ current
$\hat{T}^{grav}_{+-}$. On the
other hand the BRST current for the $SL(2,R)$ topological gauge model is based
on the $SL(2,R)$
Kac-Moody currents ${J^{gauge}}_+$. Thus it would be far from obvious whether
these two
BRST charges really give identical physical spaces. Conversely if they are
truly equivalent to each
other, then it would suggest some relations linking the Virasoro algebra and
the $SL(2,R)$ Kac-Moody
algebra. Actually it has been already known that the gauged
Wess-Zumino-Novikov-Witten (WZNW) model
with $SL(2,R)$ gauge group is reduced to the so-called gravitational WZNW model
by means of the Hamiltonian
reduction. \Ref\BO{M. Bershadsky and H. Ooguri, Commun. Math. Phys. 126 (1989)
49.}
\Ref\Pb{A. M. Polyakov, Int. Jour. Mod. Phys. A5 (1990) 833, also in ; Physics
and Mathematics of
Strings, eds. L. Brink, A. M. Polyakov and D. Friedan (World Scientific,
Singapore, 1990) p.1.}
In this respect it seems to be very interesting to clarify the equivalence in
the algebraic point of
view, though we are not going to touch on this subject in this paper.

4.1 The topological gauge theory in the "light-cone" gauge

In this section we would like to consider the equivalence from a different
point of view without
concerning the physical states themselves. We should remember that the gauge
conditions imposed to the
gravity theory and to the gauge theory differed from each other. Especially it
should be noted that
the $A_- ^a=0$ gauge (3.7) seems to be improper, if we want to regard the
topological gauge theory
as the gravitational theory. In order to compare the quantum theories started
from the actions (2.1)
and (3.1) directly, therefore, we shall examine the "light-cone" gauge which is
defined here by
$$
\eqalign{
e_+ ^+ &= e_- ^- =1, \cr
e_- ^+ &= 0, \cr
e_+ ^- &=  h_{++}, \cr
}\eqno(4.1)
$$
to also the $SL(2,R)$ topological gauge theory. These gauge conditions give us
$e=e_+ ^+ e_- ^- - e_+
^- e_- ^+ =1$, which should be compared with the previous gauge (3.7). The
light-cone gauge
$g_{--}=0$, $g_{+-}=1$ and $g_{++}=2h_{++}$ also follow from these conditions.

After fixing the gauge transformations (3.6) by the gauge conditions (4.1), we
obtain the quantum
action
$$
\eqalign{
S=\int d^2x \bigg\{
&\phi^+ \left( -\omega_+ - \partial_- h_{++} + \omega_- h_{++}\right)
          \cr
&- \phi^-\omega_- + \phi \left( \partial_+ \omega_- - \partial_- \omega_+
-1\right)
 +\tilde{b}_{++} \left( \partial_- \tilde{c}^+ + \omega_- \tilde{c}^+ \right)
          \cr
&+\tilde{b}_{-+} \left( \partial_+ \tilde{c}^+ + \omega_+ \tilde{c}^+ -
\tilde{c}\right)
 +\tilde{b}_{+-} \left( \partial_- \tilde{c}   - \omega_- \tilde{c}   +
\tilde{c}\right)
\bigg\}     \cr
}\eqno(4.2)
$$
through the BRST procedure. The BRST transformations leaving this action
invariant are found to be
$$
\eqalign{
&\delta^B h_{++}         =\partial_+\tilde{c}^- - \omega_+\tilde{c}^- +
h_{++}\tilde{c},  \cr
&\delta^B \omega_+       =\partial_+\tilde{c}   + \tilde{c}^- -
h_{++}\tilde{c}^+ ,       \cr
&\delta^B \omega_-       =\partial_- \tilde{c} - \tilde{c}^+ ,      \cr
&\delta^B \phi^+         =\phi  \tilde{c}^+ - \phi^+\tilde{c}  ,    \cr
&\delta^B \phi^-         =\phi^-\tilde{c}   - \phi  \tilde{c}^-,    \cr
&\delta^B \phi           =\phi^+\tilde{c}^- - \phi^-\tilde{c}^+,    \cr
&\delta^B \tilde{c}^+    =\tilde{c}^+\tilde{c}  ,   \cr
&\delta^B \tilde{c}^-    =\tilde{c}\tilde{c}^-  ,   \cr
&\delta^B \tilde{c}      =\tilde{c}^-\tilde{c}^+,   \cr
&\delta^B \tilde{b}_{++} =-\partial_+\phi^- + \omega_+\phi^- - \phi h_{++} +
\tilde{b}_{++}\tilde{c}, \cr
&\delta^B \tilde{b}_{-+} =-\partial_-\phi^- - \omega_-\phi^- + \phi -
\tilde{b}_{-+}\tilde{c},        \cr
&\delta^B \tilde{b}_{+-} =-\partial_+\phi^+ - \omega_+\phi^+ + \phi +
\tilde{b}_{+-}\tilde{c}.        \cr
}\eqno(4.3) $$
However, if we redefine the variables as
$$
\eqalign{
&{\phi'} ^-      \equiv \phi^- + \partial_+\phi - h_{++}\phi^+
                        + \tilde{b}_{+-}\tilde{c}^- -
\tilde{b}_{++}\tilde{c}^+,   \cr
&{\phi'} ^+      \equiv \phi^+ - \partial_-\phi - \tilde{b}_{-+}\tilde{c}^+,
   \cr
&\omega' _+      \equiv \omega_+ + \partial_-h_{++},      \cr
&\tilde{b}'      \equiv \tilde{b}_{-+} - \tilde{b}_{+-},  \cr
&\tilde{c}'      \equiv \tilde{c} + \partial_-\tilde{c}^-,\cr
&\tilde{b}       \equiv \tilde{b}_{-+},                   \cr
&\tilde{c}^{'-}  \equiv \tilde{c}^- - h_{++}\tilde{c}^+,  \cr
&\tilde{b}'_{++} \equiv \tilde{b}_{++} + h_{++}\tilde{b}, \cr
}\eqno(4.4)
$$
where it should be noted that no Jacobians appear through these redefinition,
then the action (4.2)
turns out to be
$$
\eqalign{
S=\int d^2x \bigg\{
&\phi \left( \partial_-^2 h_{++} + 1 \right) +
 \tilde{b}'_{++}\partial_-\tilde{c}^+ +
 \tilde{b} \left( \partial_+\tilde{c}^+ + \partial_-\tilde{c}^{'-} \right) \cr
&-\phi^{'-}\omega'_- - \phi^{'+}\omega'_+ + \tilde{b}'\tilde{c}'
\bigg\}.   \cr
}\eqno(4.5)
$$
Here we may take away the fields $\phi'^{\pm}$, $\omega'_{\pm}$, $\tilde{b}'$
and $\tilde{c}'$,
since they are non-dynamical and are completely decoupled from the others.
Taking account of this
we see the action (4.5) is just same as the quantum action (2.12) for
Jackiw-Teitelboim's model in the
light-cone gauge.

Next we need to examine also the BRST charge which determines the physical
space. Through the
redefinition of the variables (4.4) the BRST transformations (4.3) will be
changed into the
followings,
$$
\eqalign{
\delta^B h_{++}          =&\partial_+\tilde{c}^{'-} +
\tilde{c}^+\partial_+h_{++} +
                           \tilde{c}^{'-}\partial_-h_{++} +
2h_{++}\partial_+\tilde{c}^+   \cr
                          &-h_{++}\left( \partial_+\tilde{c}^+ +
\partial_-\tilde{c}^{'-} +
                           h_{++}\partial_-\tilde{c}^+ \right),
           \cr
\delta^B \phi            =&\tilde{c}^+ \partial_+\phi +
\tilde{c}^{'-}\partial_-\phi,      \cr
\delta^B \tilde{c}^+     =&\tilde{c}^+   \partial_+\tilde{c}^+ +
                           \tilde{c}^{'-}\partial_-\tilde{c}^+
           \cr
                          &-\tilde{c}^+ \left(\partial_+\tilde{c}^+ +
\partial_-\tilde{c}^{'-}
                           + h_{++}\partial_-\tilde{c}^+ \right) -
                           \tilde{c}^{'-}\partial_-\tilde{c}^+,
           \cr
\delta^B \tilde{c}^{'-}  =&\tilde{c}^+   \partial_+\tilde{c}^{'-} +
                           \tilde{c}^{'-}\partial_-\tilde{c}^{-'}
           \cr
                          &+h_{++}\tilde{c}^+\left(\partial_+\tilde{c}^+ +
\partial_-\tilde{c}^{'-}
                           + h_{++}\partial_-\tilde{c}^+ \right) +
                           h_{++}\tilde{c}^{'-}\partial_-\tilde{c}^+,
           \cr
\delta^B \tilde{b}'_{++} =&T^N_{++} + T^{N,gh}_{++}
                           +(h_{++})^2 \partial_-^2\phi
           \cr
                          &+ h_{++}\left( \tilde{c}^{'-} + h_{++}\tilde{c}^+
\right)
                           \partial_- \tilde{b}
                           -\left( \tilde{c}^{'-} + h_{++}\tilde{c}^+\right)
                            \left(\partial_+\tilde{b} +
\partial_-\tilde{b}'_{++}\right),  \cr
\delta^B \tilde{b}       =&T^N_{+-} + T^{N,gh}_{+-}
                           +h_{++} \partial_-^2\phi
                           + h_{++} \tilde{c}^+ \partial_- \tilde{b}
                           -\tilde{c}^+\left(\partial_+\tilde{b}
+\partial_-\tilde{b}'_{++}\right) ,
                           \cr
}\eqno(4.6)
$$
where $\hat{T}_{++}$ and $\hat{T}_{+-}$ are the same combinations of the fields
as those defined by
(2.27) and (2.29). $T^{N,gh}_{++}$ and $T^{N,gh}_{+-}$ are newly defined as
$$
\eqalign{
T^{N,gh}_{++} &= \tilde{c}^+ \partial_+\tilde{b}'_{++} +
2\partial_+\tilde{c}^+\tilde{b}'_{++}
                 + \tilde{c}^{'-}\partial_-\tilde{b}'_{++} ,            \cr
T^{N,gh}_{+-} &= \tilde{c}^+ \partial_+\tilde{b} +
\tilde{c}^{'-}\partial_-\tilde{b} .  \cr
}\eqno(4.7)
$$
If these are compared with the BRST transformations obtained previously in
(2.13), then we notice the
disagreement. However the deviation from (2.13) are found to disappear
on-shell. We may suppose that
this on-shell equivalence between the BRST transformations (4.6) and (2.13) is
caused by the
reducibility between the gauge transformations and the general coordinate
transformations. Anyhow we
may easily show that the BRST currents corresponding to the BRST
transformations (4.6) just coincide
with the former ones;
$$
\eqalign{
{J^{BRST}}_+ &= c^+ \left( T^N_{++} +{1 \over 2} T^{N,gh}_{++} \right)
              + c^- \left( T^N_{+-} +{1 \over 2} T^{N,gh}_{+-} \right), \cr
{J^{BRST}}_- &= c^+ T^N_{+-} .                                          \cr
}\eqno(4.8)
$$
Thus not only the quantum action but also the BRST charge are identical to
those of the
Jackiw-Teitelboim's model completely. Therefore we may conclude that the two
models are actually
equivalent in the quantum sense as far as we use the light-cone gauge.

4.2 The gauge a la Polyakov

It is said generally that the physical spectrum is independent of the choice of
the gauge condition.
If we can apply this argument to our case, then it would mean that we have
already proven the
quantum equivalence between the two models, since we have seen the equivalence
in the light-cone
gauge. However this seems to be too naive. Because the global modes in the
physical spectrum could
depend on the gauge choice, though the local dynamics is, of course, gauge
independent. on the other
hand Jackiw-Teitelboim's model as well as the topological gauge model is free
from any local
freedom. Therefore the physical spaces may be spanned by only the global modes.
Thus, in turn, the
equivalence between the light-cone gauge and the $A_-^a=0$ gauge seems to be
impotant to explore.

In this subsection we would like to re-examine the $SL(2,R)$ topological gauge
theory in another
type of gauge. The gauge conditions are rather similar to the $A_-^a=0$ gauge
and are given by
$$
\eqalign{
e_- ^- &=1 , \cr
e_- ^+ &= \omega_- =0 . \cr
}\eqno(4.9)
$$
This gauge has been considered by Polyakov \refmark{\Pb} so as to show the
reduction from the gauged
$SL(2,R)$ WZNW model to the gravitational WZNW model. In this gauge the "volume
form" is given by
$e=e_+ ^+$, therefore it may take any value irrespective to positive or
negative unlike the gauges
previously considered. Therefore the difference between $e$ and $|e|$ would
become sensible.

Following the BRST procedure the quantum action in this gauge is found to be
$$
\eqalign{
S= \int d^2x \bigg\{
&-\phi^+(\omega_+ + \partial_-h_{++}) - \phi^-\partial_-e_+^+ + \phi(e_+^+ -
\partial_-\omega_+) \cr
&+\tilde{b}_{++}\partial_-\tilde{c}^+ + \tilde{b}_{+-}(\partial_-\tilde{c}^- +
\tilde{c})
+\tilde{b}_+(\partial_-\tilde{c} - \tilde{c}^+) \bigg\} .    \cr
}\eqno(4.10)
$$
As is easily seen this action is invariant under the same BRST transformations
given in (3.9). Also
the BRST current is found to be identical to the BRST current obtained in the
$A_-^a =0$ gauge (3.11).
Thus the quantum theory in Polyakov's gauge looks very similar to the one in
the $A_-^a =0$ gauge.
Actually we may show they are completely equivalent in the quantum sense by
performing some field
redefinitions.

However we can reduce the number of the dynamical variables in this gauge as
follows. If we introduce
new variables as
$$
\eqalign{
&e_+^{'+}        = e_+^+ - \partial_-\omega_+, \cr
&\omega'_+       = \omega_+ + \partial_-h_{++}, \cr
&\phi^{'+}       = \phi^+ + \partial_-^2\phi^-, \cr
&\phi'           = \phi + \partial_-\phi^-, \cr
&\tilde{c}^{'+}  = \tilde{c}^+ - \partial_-\tilde{c}, \cr
&\tilde{c}'      = \tilde{c} + \partial_-\tilde{c}^-, \cr
&\tilde{b}'_+    = \tilde{b}_+ + \partial_-\tilde{b}_{++}, \cr
&\tilde{b}'_{+-} = \tilde{b}_{+-} + \partial_-^2\tilde{b}_{++}, \cr
}\eqno(4.11)
$$
then we may rewrite the action (4.10) into
$$
S=\int d^2x \bigg\{ \phi^-\partial_-^3 h_{++} -
\tilde{b}_{++}\partial_-^3\tilde{c}^- \bigg\},
\eqno(4.12)
$$
where we have eliminated the non-dynamical fields and have replaced $e_+^-$ to
$h_{++}$.
After some calculation the BRST current also turns out to be
$$
{J^{BRST}}_+ =
\left( -\partial_-^2 \tilde{c}^- + \partial_-\tilde{c}^-\partial_-
-\tilde{c}^-\partial_-^2 \right)
\left( {J^{grav}}_+^- + {1 \over 2}{J^{gh}}_+^- \right),
\eqno(4.13)
$$
where the currents ${J^{grav}}_+^-$ and ${J^{gh}}_+^-$ are given by
$$
\eqalign{
{J^{grav}}_+^- &=\partial_+\phi^- + \partial_-h_{++}\phi^-
-h_{++}\partial_-\phi^-,\cr
{J^{gh}}_+^-   &=\tilde{b}_{++}\partial_-\tilde{c}^- -
\partial_-\tilde{b}_{++}\tilde{c}^-. \cr
}\eqno(4.14)
$$

Now we should compare the action (4.12) and the BRST charge given by (4.12)
with those obtained in
the light-cone gauge (2.19) and (2.20). However we may note that the equation
of motion for the
gravitational field $h_{++}$
$$
\partial^3 h_{++} =0
\eqno(4.15)
$$
has lost the information of the cosmological constant contrary to $\partial^2
h_{++} + \Lambda=0$ in
(2.21). Therefore we may expect naively that the quantum theory in Polyakov's
gauge or in the $A_-^a
=0$ gauge has a slightly different physical space from that in the light-cone
gauge. Of course our
analysis is, however, far from complete to show the inequivalence.
Investigation in more detail on
the BRST cohomologies will be required.

\chapter{Topological algebra in Jackiw-Teitelboim's model}

In section 3 we have found that the twisted N=2 SCA is indeed realized in the
$SL(2,R)$ topological
gauge theory quantized in the $A_-^a=0$ gauge as was shown in (3.16). However
it seems to be hard to
expect that Jackiw-Teitelboim's model, which has been shown to be equivalent to
the $SL(2,R)$
topological gauge theory quantized in the light-cone gauge, also has such a
twisted N=2 SCA. Because
the $SL(2,R)$ symmetry seems to have been completely lost in the
Jackiw-Teitelboim's model. Moreover
$\bar{G}_{++}$ defined in (3.15) is the generator for the spin-1 supersymmetry,
which may be readily
seen from the action (3.8). On the other hand there seems to be no
supersymmetry realized in the
action (2.19).

However we may find out another kind of the twisted N=2 SCA in the light-cone
gauge. Define the
generators for the twisted N=2 SCA in terms of the fields introduced in (2.22)
as
$$
\eqalign{
T_{++}       = &\hat{T}^{grav}_{++} + \hat{T}^{gh}_{++},  \cr
G_+          = &\hat{c}^+ \hat{T}^{grav}_{++} + \hat{c}^+\partial_+ \hat{c}^+
\hat{b}_{++} +
                \hat{c}^- \hat{c}^+ \partial_+ \hat{b}  \cr
               &+\partial_+\left(\hat{c}^+ b \hat{c}^- \right)
                -\partial_+\left[ c^+\left( {1 \over 2}\hat{h}_+ -
\partial_+\hat{\phi} \right)\right],  \cr
\bar{G}_{++} = &\hat{b}_{++}, \cr
I_+          = &-\hat{b}_{++}\hat{c}^+ - \hat{b}\hat{c}^- -
\hat{h}_{++}\hat{\phi}^+ +
                \left( {1 \over 2}\hat{h}_+ - \partial_+\hat{\phi} \right),
\cr
}\eqno(5.1)
$$
where $\hat{T}^{grav}_{++}$ and $\hat{T}^{gh}_{++}$ are given in (2.27) and in
(2.29). Then we can
verify that these currents indeed form a twisted N=2 SCA like (3.16) but with
the central charge
$\hat{c}=0$ by using the O.P.E.'s given in (2.25) and (2.26). Here it should be
noted that the spin-1
fermionic current $G_+$ is not exactly same as the BRST current $J^{BRST}_+$ in
(2.20). If, however,
we set $\hat{T}^{grav}_{+-}=0$, then the BRST current is reduced to $G_+$ up to
the total divergent
terms. Actually we may see that the physical states $|phys>$ satisfy
$\hat{T}^{grav}_{+-}|phys>=0$.
Therefore $G_+$ may be play the same role as the BRST current effectively on
the physical space. We
also note that $I_+$ is not the ghost number current unlike $I_+$ in (3.15).

Recently Fujikawa and Suzuki
\Ref\FS{K. Fujikawa and H. Suzuki, Nucl. Phys. Nucl. Phys. B361 (1991) 539.}
 have found that the two dimensional gravity coupled to $c=-2$ matter has also
the twisted N=2 SCA
with the central charge $\hat{c}=0$. In this case also the spin-1 fermionic
current is given by the
BRST current up to total derivatives. Their analysis, however, has been done in
the conformal gauge.
It would be natural to expect that we may find out such a twisted N=2 SCA for
the theory quantized in
the light-cone gauge also. This is indeed the case and the SCA in the
light-cone gauge will be found
to have quite similar structure to the algebra formed by the generators in
(5.1).
\Ref\Y{K. Yamada, preprint DPKU-9208 (1992).}
The readers should refer the ref.[\Y] for more details. It seems to be also
interesting to examine
the topological algebra for Jackiw-Teitelboim's model quantized in the
conformal gauge.

\ack

I would like to thank Prof. N. Nakanishi, Prof. N. Kawamoto, M. Abe, H.
Kunitomo, I. Oda and H. Ooguri for useful
conversations, and K. Yamada for explaining his work in ref.[\Y]. I am
especially grateful to K. Itasaka for
 his collaboration in the early stage of this work. Finally, I would like to
thank Prof. T. Suzuki and Prof. J.
Kubo for discussions and encouragement.

\refout
\bye